\documentclass[fleqn,12pt,twoside]{article}
\usepackage[headings]{espcrc1}

\usepackage{graphicx}
\usepackage[figuresright]{rotating}

\newcommand{\AmS}{{\protect\the\textfont2
  A\kern-.1667em\lower.5ex\hbox{M}\kern-.125emS}}

 \newcommand\la{\langle}
 \newcommand\ra{\rangle}
 \newcommand\beq{\begin{equation}}
 
 \newcommand\eeq{\end{equation}}
 \newcommand\beqn{\begin{eqnarray}}
 \newcommand\eeqn{\end{eqnarray}}

\def\mb{\,\mbox{mb}}
\def\fm{\,\mbox{fm}}
\def\GeV{\,\mbox{GeV}}
\def\MeV{\,\mbox{MeV}}

\def\Pom{{\bf I\!P}}
\def\Reg{{\bf I\!R}}
\def\lsim{\mathrel{\rlap{\lower4pt\hbox{\hskip1pt$\sim$}}
    \raise1pt\hbox{$<$}}}         
\def\gsim{\mathrel{\rlap{\lower4pt\hbox{\hskip1pt$\sim$}}
    \raise1pt\hbox{$>$}}}         

\title{Glue drops inside hadrons}

\author{B.Z.~Kopeliovich\address[vina]{Universidad
Tecnica Federico Santa Maria, Valparaiso, Chile}\address[hd]{Institut
f\"ur Theoretische Physik der Universit\"at,
Heidelberg, Germany},
B.~Povh\address[mpi]{Max-Planck-Institut f\"ur Kernphysik, 69029
Heidelberg, Germany},
Ivan~Schmidt\addressmark[vina]}

\begin{document}

\maketitle

\begin{abstract}

We present experimental evidences for the existence of a semi-hard scale
in light hadrons.  This includes the suppression of gluon radiation that
is seen in high mass hadron diffraction; the weak energy dependence of
hadronic total cross sections; the small value of the Pomeron trajectory
slope measured in photoproduction of $J/\Psi$; the weakness of gluon
shadowing in nuclei; shortage of gluons in the proton revealed by an
unusual behavior of the proton structure function in the soft limit, and
the enhanced intrinsic transverse momentum of quarks and gluons, which
considerably exceeds the inverse hadronic size. All these observations
suggest that gluons in hadrons are located within spots of a small size
relative to the confinement radius.

\end{abstract}



\section{Introduction}

 There is growing theoretical and experimental support leading towards the
existence of a non-perturbative scale smaller than the usual
$1/\Lambda_{QCD} \sim 1\fm$, and which is related to the gluonic
degrees of freedom. First, an analysis of hadronic matrix elements
of the gluonic contribution to the energy momentum tensor, using
the QCD sum rules approach, gives a value of 0.3 fm for the radius
of the corresponding form factor \cite{1}. From the lattice side,
numerical simulations of the gluon two point correlation function
turn out a value of also $0.2-0.3\fm$ for the correlation length
\cite{2}, and the energy of the QCD string appears concentrated in
a tube of radius $0.3\fm$ in the transverse direction \cite{3}. On
the other side, it has been shown that the instanton radius peaks
approximately at $1/3\fm$ \cite{4}.  Furthermore, high statistics
data for diffractive gluon bremsstrahlung in hadronic collisions
is difficult to explain unless gluons in the proton have
transverse momenta as high as about $0.7\GeV$ \cite{kst2}. This
has been confirmed by studies of diffractive parton distributions,
which concluded that they have a rather small transverse size
\cite{5}. What actually happens is that the smallness of the gluon
clouds slows down Gribov diffusion of the gluons in transverse
plane, and this results in a small slope of the Pomeron trajectory
in hard reactions \cite{kp}, in agreement with data. More
arguments in favor of small gluonic spots coming from DIS can be
found in \cite{sz}. Some of these results have been corroborated
by recent studies of the spatial distribution of gluons in the
transverse direction at small $x$ \cite{6}. Here we overview the
available experimental evidences for the presence of a semihard
scale in hadronic structure.

\section{Why gluon radiation is suppressed}
\label{diff}

If gluons in hadrons are located within small spots of radius
$r_0$, they have enlarged transverse momenta $q_T\sim 1/r_0$. Such
gluons cannot be resolved by soft interactions and be shaken off,
which means that the bremsstrahlung cross section should be
suppressed compared to perturbative estimates.

However, in the case of soft inelastic collisions followed by
multiparticle production the events with or without gluon
radiation look alike. In both cases the produced particles build a
plateau in rapidity, and then it is difficult to find any definite
signature of the radiated gluons.

Diffraction offers an exclusive possibility to identify gluon
radiation. A high-energy hadron can dissociate diffractively
either via excitation of the valence quark skeleton, or by
radiating gluons. These two mechanisms are characterized by
different dependence on the effective mass, $M_X$, of the
excitation,
 \beqn
\frac{d\sigma(hp\to X p)}{dM_X^2} =
\left\{
\begin{array}{lc}%
\frac{1}{M_X^3} & {\rm excitation\ of\ the\ quark\ skeleton}\\
\frac{1}{M_X^2} & {\rm diffractive\ gluon\ bremsstrahlung}
\end{array}
\right.
\label{100}
 \eeqn
 The $M_X$-dependence at large $M_X$ correlates with the spin of the
slowest particle produced in the excitation. Only a vector
particle, i.e. a gluon, can provide the $1/M_X^2$ dependence.

\begin{figure}[h]
\begin{center}
 \includegraphics[width=10cm]{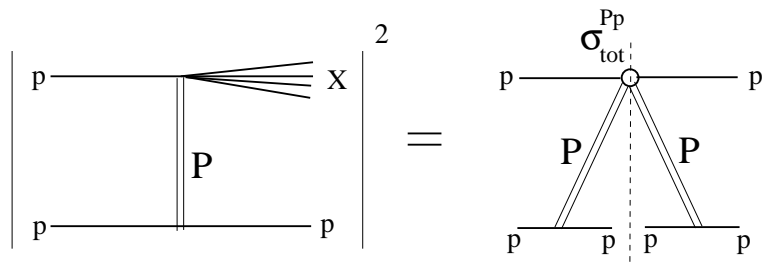}
 \end{center}
 {\caption[Delta]{The cross section of diffractive excitation of a proton
expressed in terms of the total Pomeron-proton cross section.}
 \label{3R}}
 \end{figure}

Thus, one can single out the cross section of diffractive gluon
radiation from the large mass tail of the $M_X$-distribution. An
analysis \cite{kklp} of diffractive data shows that gluon
radiation is amazingly weak. In order to understand that one can
interpret diffraction in terms of the Pomeron-proton total cross
section, as is shown in Fig.~\ref{3R}.
 If we treat the Pomeron like a gluonic dipole, one may expect a cross
section $9/4$ times larger than for a $\bar qq$ dipole. Comparing with the
pion-proton cross section, say $25\mb$, one arrives at the estimate of
about $50\mb$. However, data suggest quite a smaller value, about $2\mb$.
A straightforward explanation for such a dramatic disagreement would be a
much smaller size of the gluonic dipole (Pomeron) compared to the
quark-antiquark dipole (pion). Thus, one concludes that gluons should be
located within small spots in the proton.

Although it is not quite rigorous, one might try to estimate the
diffractive radiation cross section perturbatively, and in this
case the result exceeds data by more than an order of magnitude.
To reduce the cross section down to the observed value one should
assume that the mean quark-gluon separation is as small as
$r_0=0.3\fm$ \cite{kst2}. With such a modified quark-gluon
light-cone distribution function, the effective triple-Pomeron
coupling has the form \cite{kst2},
 \beq
G_{3\Pom}(0) \equiv (1-x_F)^{\alpha_\Pom(0)}\,
\left.\frac{d\sigma_{sd}(pp\to pX)}
{dx_F\,dp_T^2}\right|_{p_T=0}=
\frac{81\alpha_s\sigma_0}
{(16\pi)^2}\,\ln\left[\frac{2(r_0^2+R_0^2)^2}
{R_0^2(2r_0^2+R_0^2)}\right]\ .
 \label{150}
 \eeq
 Here we assume that $1\gg 1-x_F\gg s_0/s$, where $s_0\sim 1\GeV^2$. The
energy dependent parameters $\sigma_0(s)$ and $R_0(s)$ controlling the
shape of the universal dipole cross section \cite{zkl} are defined in
\cite{kst2}. With $r_0=0.3\fm$, the triple-Pomeron coupling
eq.~(\ref{150}) agrees with the result of the triple-Regge analysis
\cite{kklp} of single diffractive data.

\section{Why energy dependence of hadronic cross sections is so weak}

It is well known that hadronic cross sections rise with energy
approximately as $s^\epsilon$, where the exponent is quite small,
$\epsilon\approx 0.1$. What is the origin of this small number? We do not
expect any small parameters in the soft regime of strong interactions.

This problem is closely related to the topic of the previous
section. In fact, the energy dependence is driven by gluon
radiation which turns out to be suppressed. Before we saw a
manifestation of this effect in diffraction, now in the total
inelastic cross section.

Without gluon radiation the geometric cross section of two hadrons
would be constant, since their transverse size is Lorentz
invariant, i.e. is energy independent. The phase space for one
gluon radiation is proportional to $\ln(s)$, so multigluon
radiation leads to powers of $\ln(s)$ in the cross section.  The
calculations performed in \cite{k3p} confirm this. The hadronic
cross section was found to have the following structure,
 \beq
\sigma_{tot}=\sigma_0 + \sigma_1\,
\left(\frac{s}{s_0}\right)^\Delta\ ,
\label{200}
 \eeq
 where $\sigma_0$ is the energy independent term related to hadronic
collisions without gluon radiation. The second term in (\ref{200})
is the contribution of gluon bremsstrahlung to the total cross
section. Here the parameter $\sigma_1$ is expected to be small due
to the smallness of the gluonic spots. Indeed, it was found in
\cite{k3p} that $\sigma_1=27\,C\,r_0^2/4$, where factor $C\approx
2.4$ is related to the behavior of the dipole-proton cross
section, calculated in Born approximation at small separations,
$\sigma(r_T)= Cr_T^2$ at $r_T\to0$.

The energy dependence of the second term in (\ref{200}) was found
to be rather steep, $\Delta=4\alpha_s/3\pi=0.17$. This exponent
seems to be too large compared to the experimentally measured
$\epsilon\approx 0.1$. There is, however, no contradiction due to
the presence of the large energy independent term in (\ref{200}).
Approximating the cross section (\ref{200}) by a simple power
dependence on energy, the effective exponent reads,
 \beq
\epsilon=\frac{\Delta}{1+\sigma_0/\sigma_1\,(s/s_0)^{-\Delta}}
\label{300}
 \eeq
 So, one should expect a growing steepness of the energy dependence for
the total cross section. One can estimate the value of $r_0$ demanding the
effective exponent to be $\epsilon\approx 0.1$ in the energy range of
fixed target experiments, say at $s\sim 1000\GeV^2$. With $\sigma_0=40\mb$
found in \cite{k3p} one gets $r_0=0.3\fm$.

Thus, the observed slow rise of the total hadronic cross sections
provides another evidence for the existence of small gluonic spots
with transverse size $r_0\sim 0.3\fm$.

One may expect a steeper energy dependence for heavier flavors.
Indeed, for $J/\Psi$-proton scattering $\sigma_0$ is
so small, that $\epsilon\approx\Delta$. Indeed, data for $J/\Psi$
photoproduction from HERA \cite{psi} show that $\epsilon\approx 0.2$. One
should be careful, however, interpreting the data within the vector
dominance model \cite{hk-vdm}, and remember that Eq.~(\ref{200}) was
derived assuming that $r_0$ is much smaller than the hadronic size,
otherwise interferences should be included.

\section{Why diffractive cone shrinks with energy slowly}

The prediction of a shrinkage of the diffraction cone has been one
of the first achievements of the Regge theory. Indeed, data show
that the elastic slope in hadronic collisions rises with energy
as, $B_{el}(s)=B_0+2\alpha_\Pom^\prime\,\ln(s/s_0)$, where
$\alpha_\Pom^\prime\approx 0.25\GeV^{-2}$. This is about four
times smaller than in binary processes mediated by other Reggeons,
$\alpha_\Reg^\prime\approx1\GeV^{-2}$. Why?

\begin{figure}[htb]
\begin{minipage}[t]{66mm}
 \includegraphics[width=65mm]{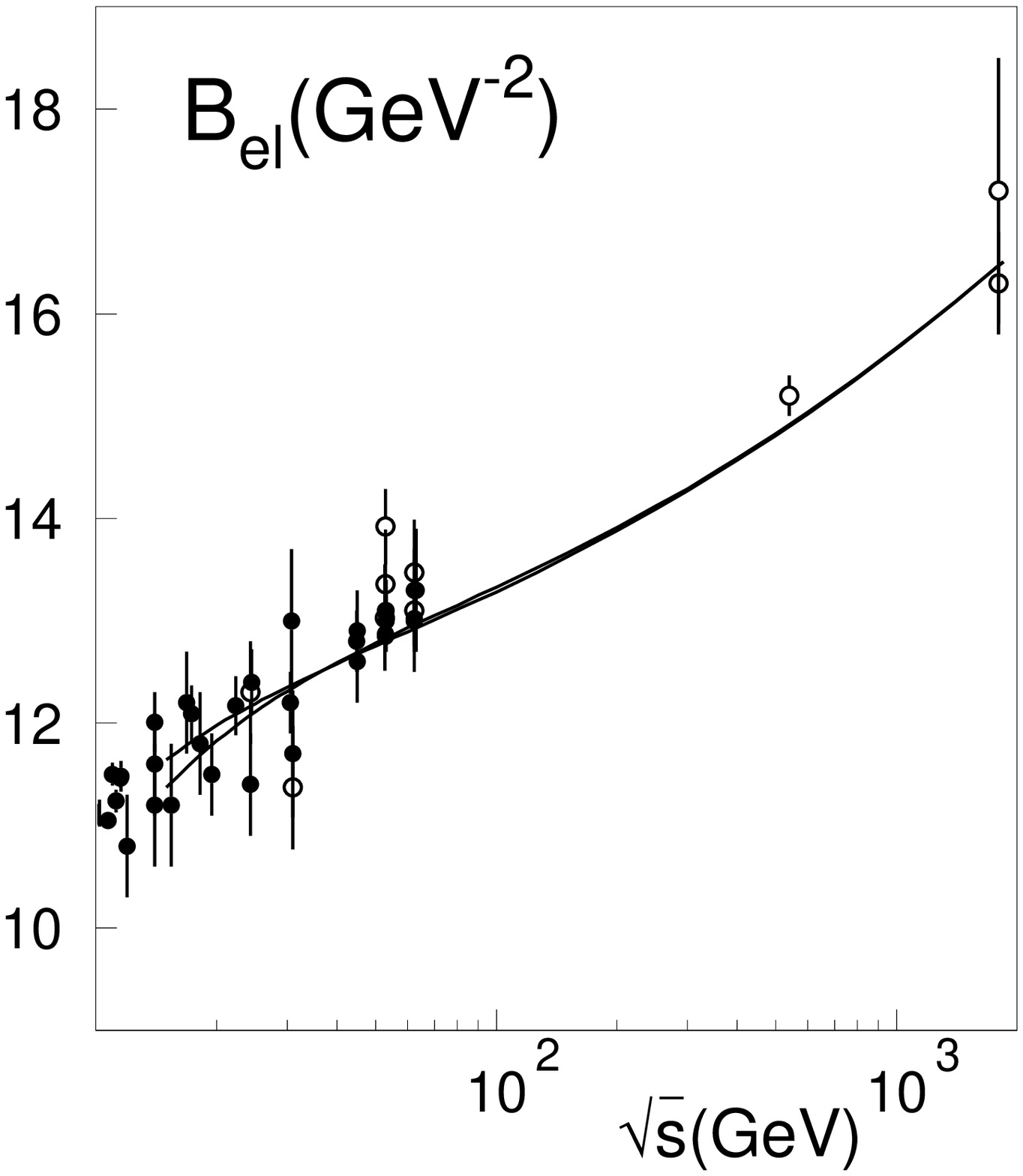}
 \caption{Elastic slope for $pp$/$\bar pp$ (bottom/upper curves)
collisions, calculated in \cite{k3p} with the small Pomeron slope
according to (\ref{500}). The references to the data can be found in 
\cite{k3p}.}
 \label{slope1}
 \end{minipage}
 \hspace{\fill}
\begin{minipage}[t]{75mm}
 \includegraphics[width=82mm]{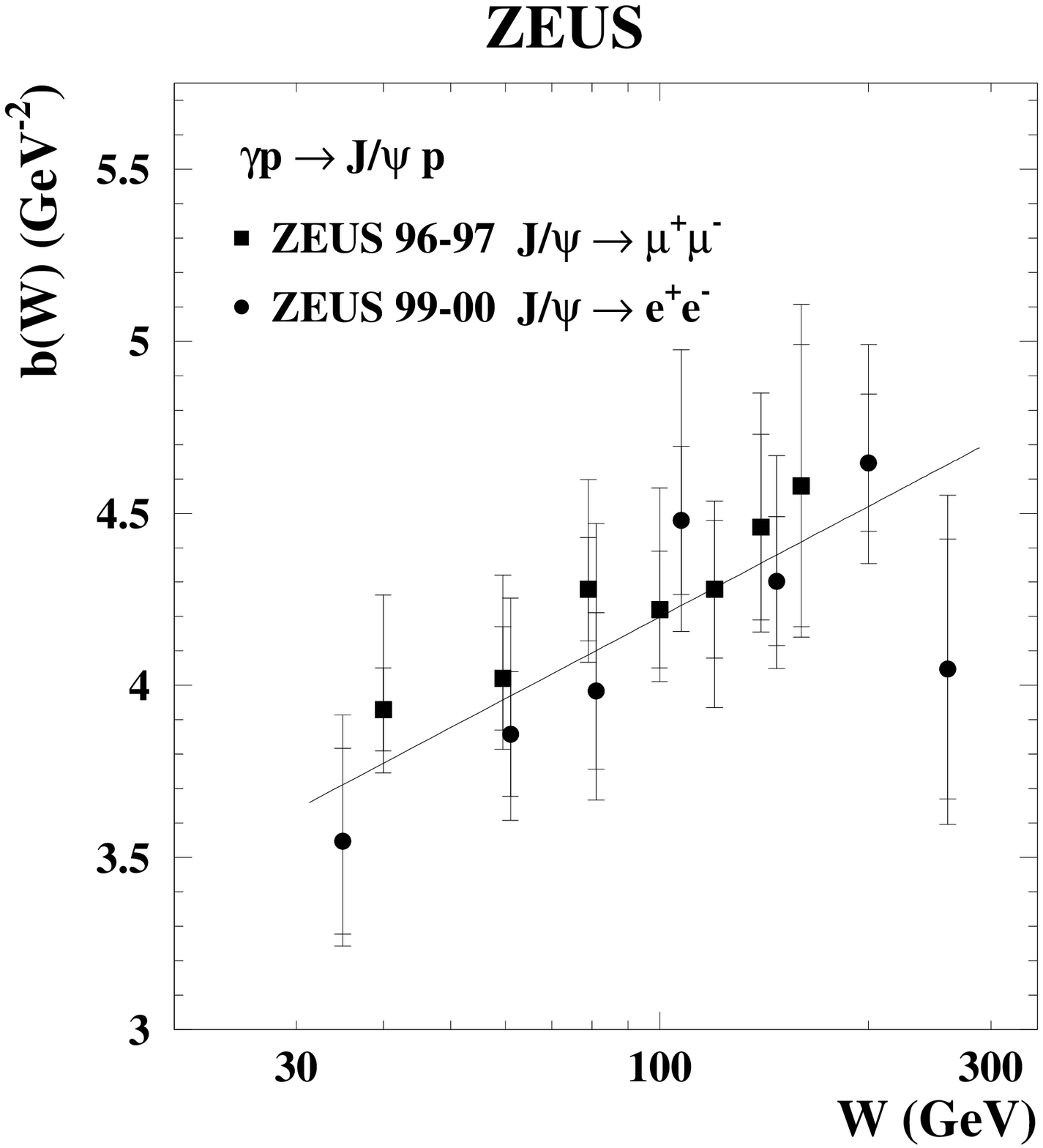}
 \caption{Elastic slope in exclusive photoproduction $\gamma+p\to
J/\Psi+p$. Data are from the ZEUS experiment \cite{psi}. The solid
curve is a fit with $\alpha_\Pom^\prime = 0.115\pm0.018\GeV^{-2}$, in good
agreement with (\ref{500}).}
 \label{slope2}
\end{minipage}
 \end{figure}

The diffractive cone shrinkage is usually related to Gribov
diffusion of gluons in the transverse plane. If each "step" in
impact parameters, $\Delta b^2=r_0^2$, is small, the diffusion
should proceed slowly. Indeed, a rather small value of the Pomeron
trajectory slope  was predicted in \cite{k3p},
 \beq
\alpha_\Pom^\prime =
{1\over2}\,\frac{dB_{el}}{d\ln(s/s_0)} =
\frac{\alpha_s}{3\pi}\,r_0^2 = 0.1\GeV^{-2}\ .
\label{500}
 \eeq
 This seems to be too small, in strong contradiction with value 
$0.25\GeV^{-2}$
known from data for the elastic slope. One may wonder, why
the same model \cite{k3p} which predicts (\ref{500}) describes well data
for elastic slope, as is demonstrated in Fig.~\ref{slope1}?
 The relatively large value of $\alpha_\Pom^\prime$, turns out to result
from unitarity saturation. Indeed, the elastic differential cross section
(actually the amplitude) Fourier transformed to impact parameter
representation (see details in \cite{k3p}) demonstrate unitarity saturation
at small impact parameters. In spite of the observed rise of the total
cross section with energy, there is no room for further growth at small
impact parameters, only the amplitudes of peripheral collisions rise with
energy. This leads to a rising with energy radius of interaction directly
related to the elastic slope. Thus, a substantial part of the observed
energy dependence of the elastic slope and of the effective
$\alpha_{eff}^\prime=0.25\GeV^{-2}$ is related to saturation of the
unitarity bound. How to disentangle the two effects?

 To get rid of unitarity corrections one can consider the interaction of a
small dipole with a proton. For example photoproduction of a heavy
quarkonium \cite{kp}, or high $Q^2$ electroproduction of $\rho$.
Then, the elastic amplitude is too small to be affected by
unitarity (absorptive) corrections, and the energy dependence of
the slope must be solely due to the rise of the gluon clouds, i.e.
Gribov diffusion. This expectation of \cite{k3p} was nicely
confirmed in elastic photoproduction $\gamma+p\to J/\Psi+p$
measured by the ZEUS experiment \cite{psi} which found
$\alpha_\Pom^\prime=0.115\pm 0.018 \GeV^{-2}$. The data and fit
are depicted in Fig.~\ref{slope2}

\section{Why gluon shadowing is so weak}

Although bound nucleons in a nucleus are rather well separated,
Lorentz-boosted into the infinite momentum frame they start
overlapping at small Bjorken $x$. This happens since Lorentz
contraction of the gluonic clouds at small $x$ is much weaker than
at large $x$. If the clouds originated from different nucleons
overlap, they may fuse reducing the gluon density at small $x$.
This phenomenon is called gluon shadowing. At first glance this
might be a considerable effect, since gluons interact stronger
than quarks. However, one should be cautious, since a similar
naive expectation of a large cross section for gluonic dipoles
failed when was confronted with data (see Sect.~\ref{diff}), and
both phenomena have common roots.

Even if gluonic clouds overlap in the longitudinal direction, one
should make sure that it also happens in the transverse plane.
This is not obvious if the clouds are small. In a nucleus the mean
number of gluonic spots overlapping at a given impact parameter
is,
 \beq
\la n\ra=\frac{3\pi}{4}\,
r_0^2\, \la T_A\ra\ ,
\label{600}
 \eeq
 where $\la T_A\ra\approx \rho_A\,{4\over3}R_A$ is the mean nuclear
thickness; $\rho_A\approx 0.16\fm^{-3}$ is the nuclear density;
and $R_A\approx 1.14\fm\times A^{1/3}$ is the nuclear radius.
Notice that $r_0$ is the mean diameter of a dipole, rather than
radius, which explains the factor $1/4$ in (\ref{600}).

According to (\ref{600}) even for the heaviest nuclei only very
few gluonic clouds have a chance to overlap at a given impact
parameter, $\la n\ra \approx 0.3$. Therefore, the longitudinal
overlap of gluons originating from different nucleons does not
lead to gluon interaction if the transverse overlap is so small.
This may substantially weaken all collective phenomena at small
$x$ including color glass condensate \cite{mv}.

\begin{figure}[htb]
\begin{minipage}[t]{80mm}
 \includegraphics[width=65mm]{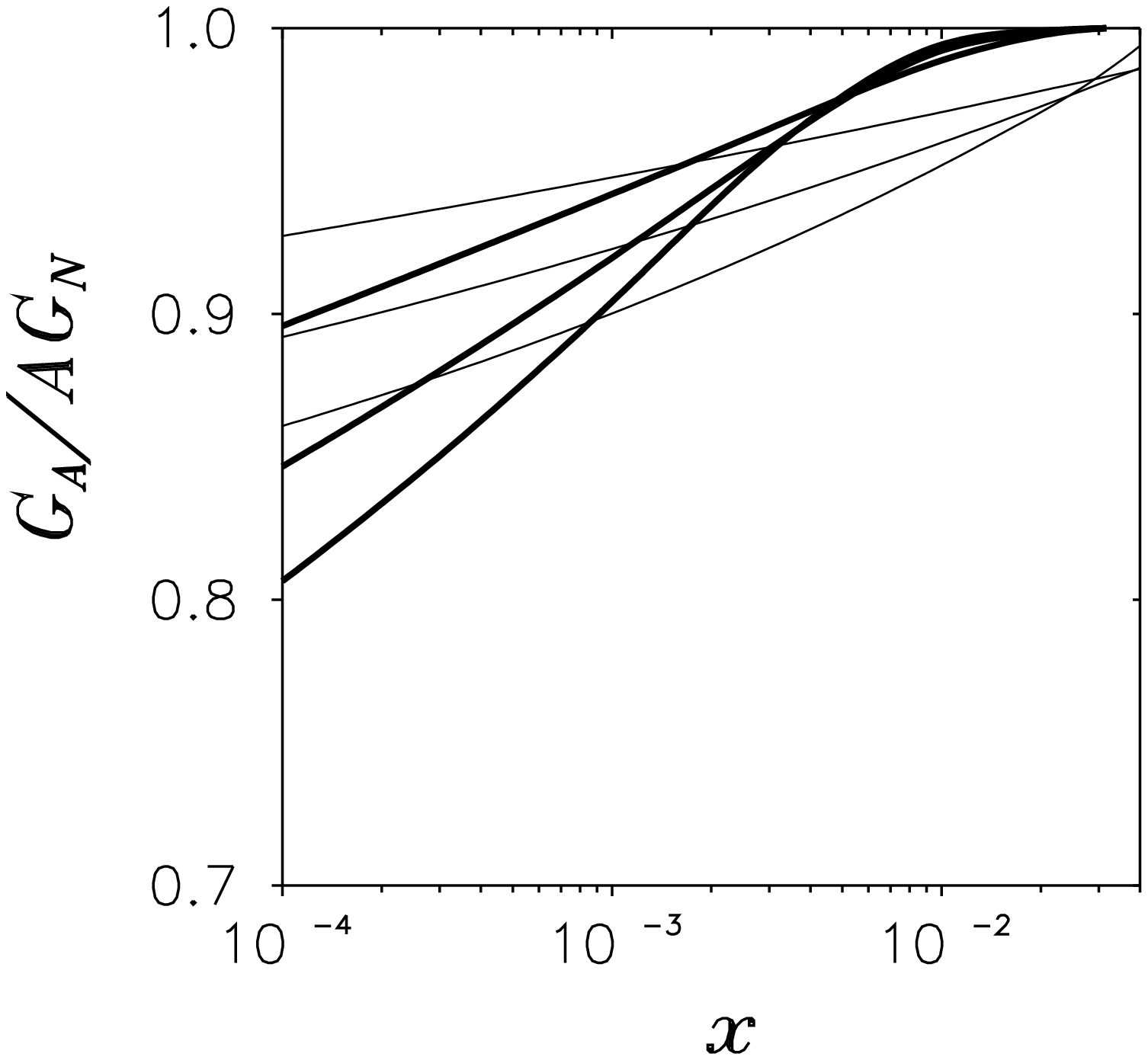}
 \caption{Predictions for gluon shadowing based on the small gluonic spot
structure \cite{kst2}. The two sets of curves are calculated by different
methods and at different scales, at $Q^2\approx 1/r_0^2$ (thin curves) and
at $Q^2=4\GeV^2$ (thick curves). Each set is calculated for carbon, iron
and led (from top to bottom).}
 \label{soft}
 \end{minipage}
 \hspace{\fill}
\begin{minipage}[t]{70mm}
 \includegraphics[width=65mm]{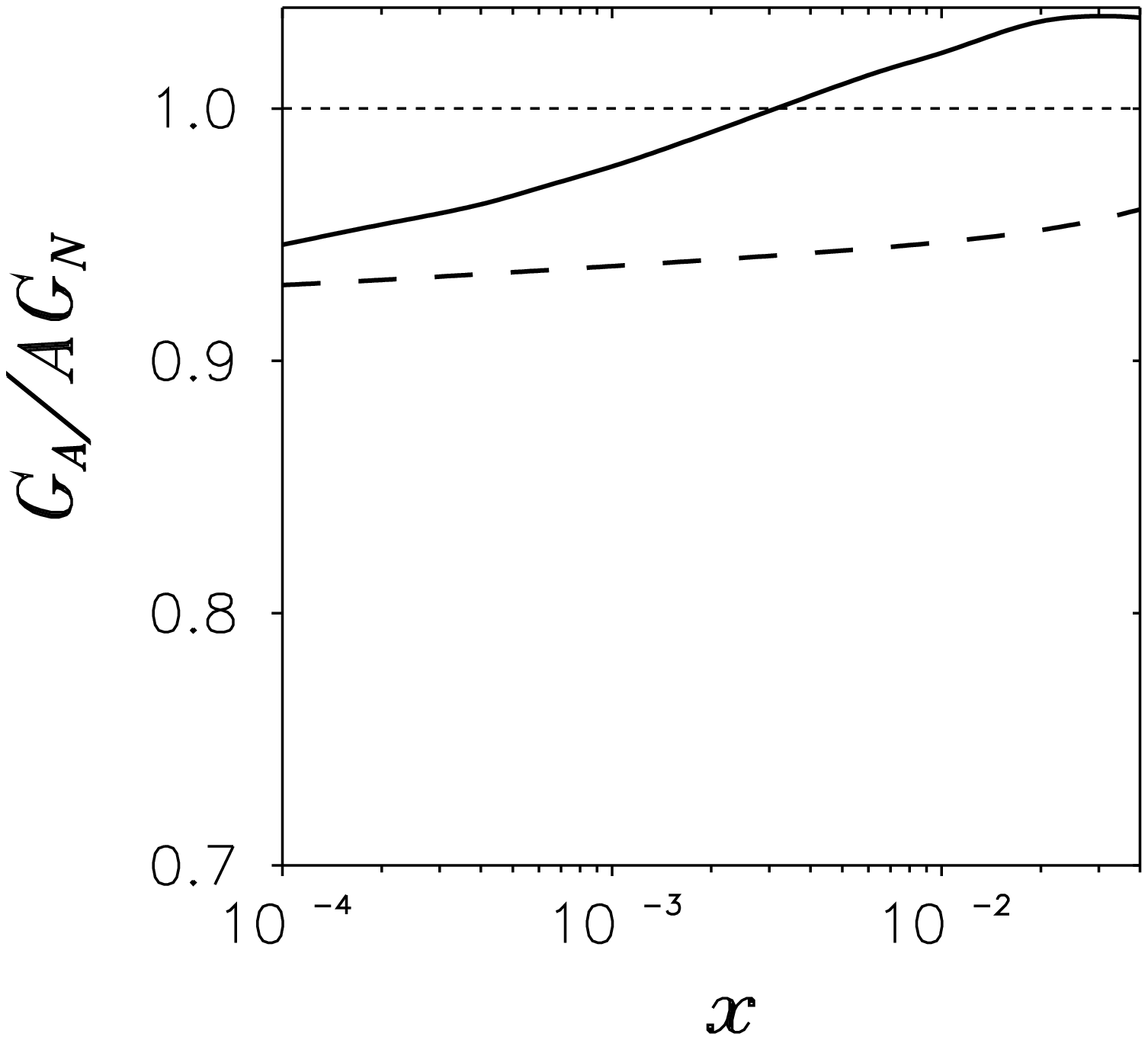}
 \caption{Gluon shadowing in calcium suggested by DGLAP analysis of
DIS data on nuclei. The solid curve is from an analysis
\cite{florian} done in the next-to-leading order approximation.
The dashed curve is the result of a leading order analysis
\cite{hkm}.}
 \label{florian}
\end{minipage}
 \end{figure}

Thus, another consequence of smallness of the gluonic spots is a
very weak gluon shadowing, predicted in \cite{kst2}.
Fig.~\ref{soft} demonstrates results of calculations performed in
\cite{kst2} by different methods and at different scales
$Q^2\approx 1/r_0^2$ and $4\GeV^2$. Even at $x$ as low as
$10^{-4}$ shadowing corrections for medium heavy nuclei are only
$10\%$.

Unfortunately, the gluon distribution function is difficult to
measure, since the main partonometer, deep-inelastic scattering
(DIS), probes directly only quarks and antiquarks. The gluon
distribution can be accessed only via $Q^2$ evolution. Although
previous attempts to single out the gluon distribution in nuclei
from DIS resulted in big uncertainties, a recent next-to-leading
order analysis \cite{florian} turned out to be quite sensitive to
gluons. Gluon shadowing extracted from data is depicted in
Fig.~\ref{florian} for calcium \cite{florian}. Dashed curve show
the results of a LO analysis \cite{hkm}. We do not compare with
the gluon shadowing suggested in \cite{eks}, since it was ad hoc.
The results of the DGLAP analyses agree quite well with
predictions \cite{kst2} within the theoretical and fitting
uncertainties. Other gluon shadowing calculations missing the
effect of small gluonic spots dramatically overestimate the
magnitude of gluon shadowing.

\begin{figure}[h]
 \begin{minipage}{7.5cm}
 \includegraphics[width=7.5cm]{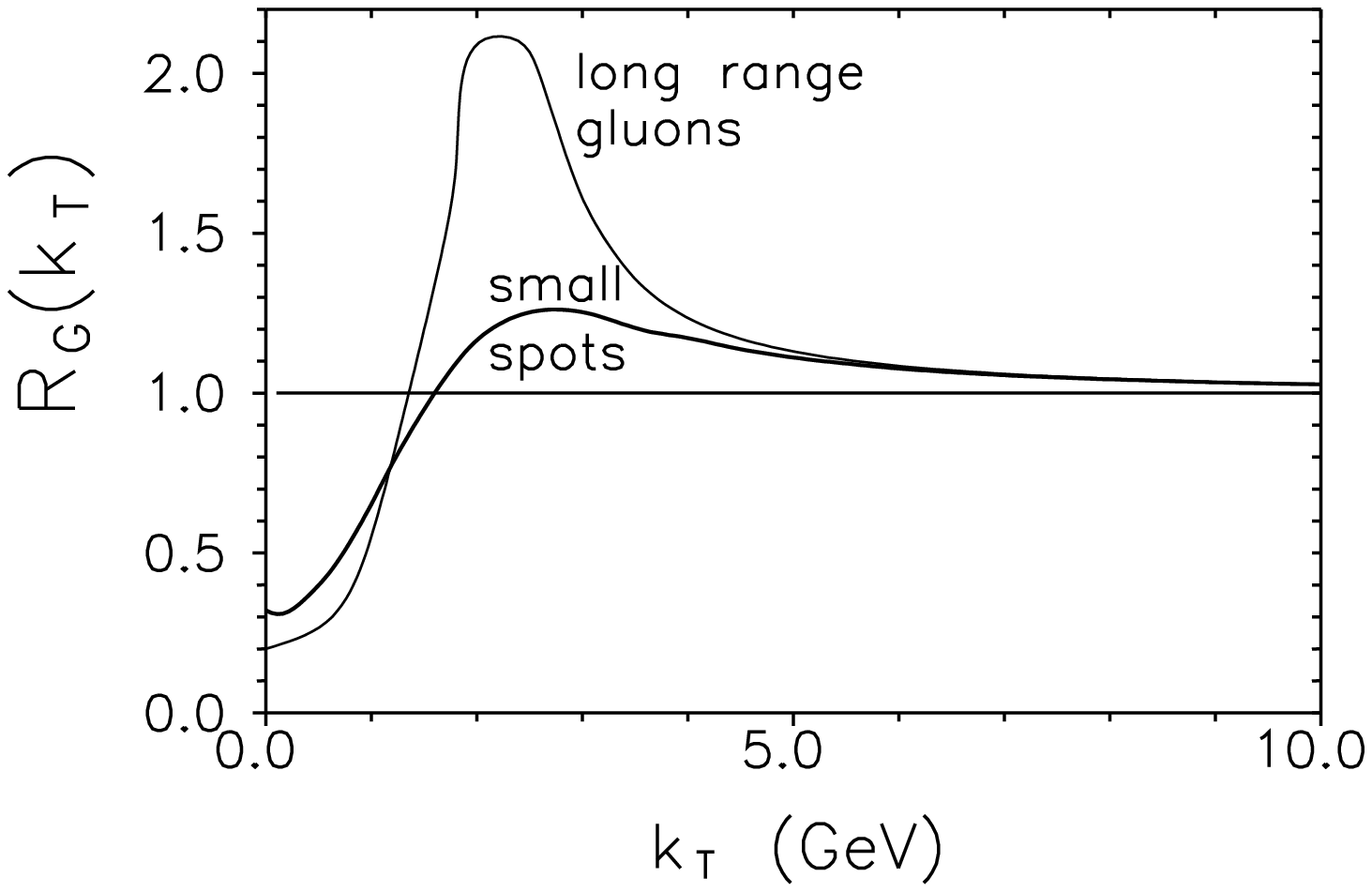}
\caption{Normalized ratios of gluon radiation cross sections in quark-nucleus
over quark-nucleon collisions. The ratios are plotted as function of gluon 
transverse momentum. Calculations \cite{kst1} are performed for lead in two cases of
long-range and short range ($r_0=0.3\fm$) gluons. In the latter case the magnitude 
of the color glass condensate effect is much smaller.}
\label{cronin}
 \end{minipage}
 \hspace{\fill}
 \begin{minipage}{7cm}
 \includegraphics[width=7cm]{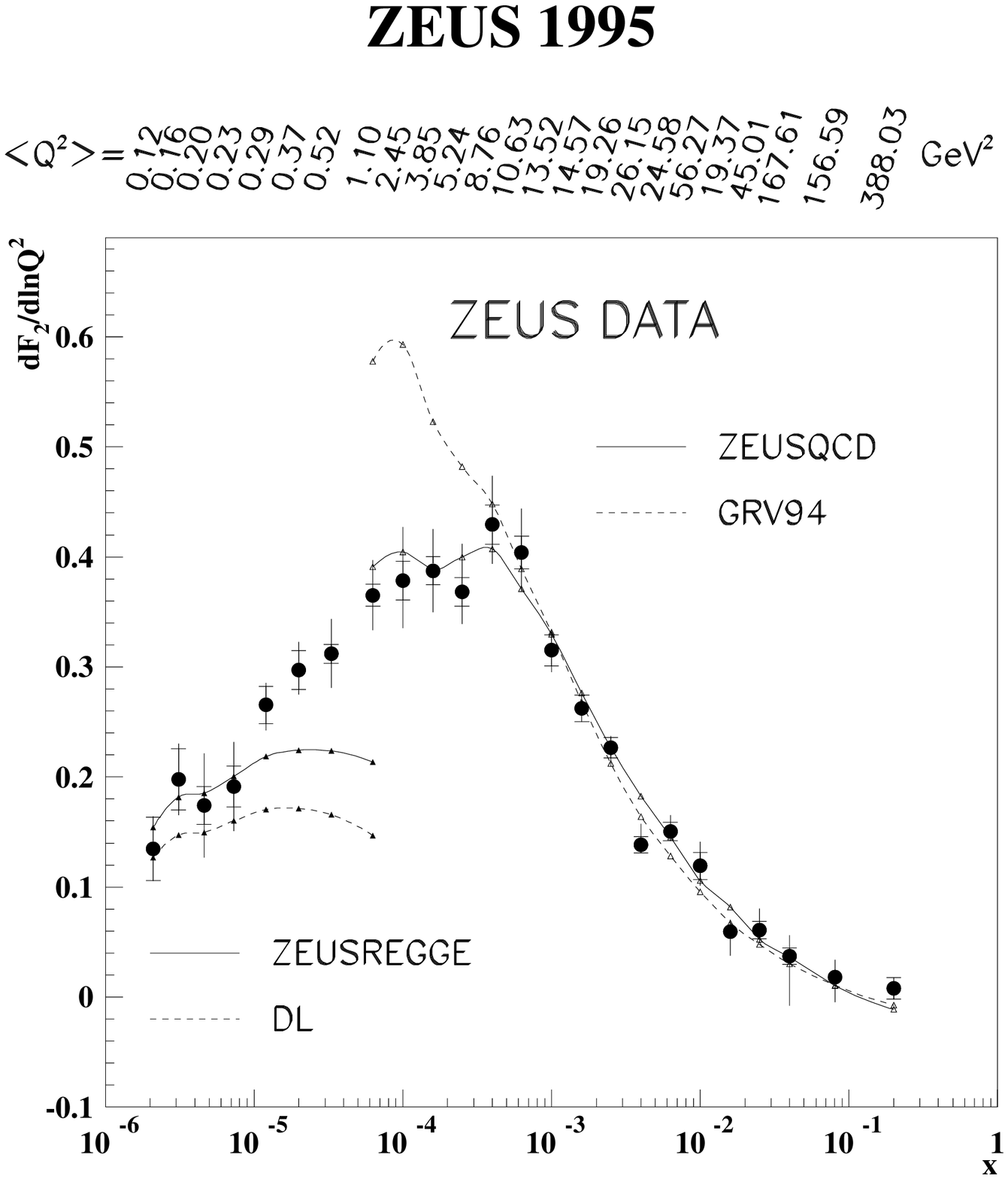}
\caption{Logarithmic $Q^2$ derivative of $F_2(x,Q^2)$ as function of $x$ 
and $Q^2$ \cite{zeus-caldwell}. The observed turn over and fall at small $Q^2$ 
indicates at shortage of gluons.}
\label{caldwell}
\end{minipage}
\end{figure}

Suppression of gluon radiation due to smallness of gluonic spots also leads to
a substantial reduction of the effect of color glass condensate. This is 
demonstrated in Fig.~\ref{cronin} where $p_T$-dependent nuclear ratios are 
calculated for long-range ($r_0\sim1/\Lambda_{QCD}$) and short range 
($r_0\sim 0.3\fm$) gluons \cite{kst1}. Gluonic spot structure of the proton also 
leads to a very weak Cronin enhancement predicted for RHIC (confirmed) and LHC
\cite{knst}.

\section{Why there is a shortage of gluons at low scale}

As far as gluons are located within small spots, it is difficult to
resolve them at low scale, $Q^2 < 4/r_0^2$. With poor resolution the
proton looks like a 3-quark system containing no gluons. At the same time,
no changes happen at higher $Q^2$ which resolve distances much smaller
than the size of gluonic spots. In this regime the gluon density is rising
toward smaller $x$ unless related values of $Q^2$ become too low to
resolve the spots. This explains the ZEUS data \cite{zeus-caldwell} depicted in
Fig.~\ref{caldwell} for logarithmic $Q^2$ derivative of the structure
function. This derivative suddenly drops at $Q^2$ below few $\GeV^2$
showing that parton distributions are frozen below this scale, no evolution
happens due to lack of gluons.

\section{Why transverse momenta of hadron constituents are so large}

If there was only one scale in hadronic structure, the charge
radius, one would expect the mean momenta of hadron constituents
to be of the order of the inverse radius, i.e. $\la
k_T\ra\sim\Lambda_{QCD}\approx 200\MeV$. This is what one would
observe with a poor resolution, insufficient for seeing the
structure of constituent quarks. However, with somewhat better
resolution one can see gluons whose Fermi motion is much more
intensive, since they are confined within small spots, $\la
k_T\ra\sim1/r_0\approx 700\MeV$. There should be manifestations of
such an enhanced Fermi motion in reactions and observables
sensitive to the primordial parton momentum.

\subsection{Sea-Gull effect}

The projectile quark and gluons lose (at least partially) coherence and
emerge from the interaction area as forward jets. Theses jets having
rather small transverse momenta hardly can be reconstructed.  However,
inclusively detected hadrons carrying fraction $z$ of the initial
quark/gluon correspondingly have same fraction of the jet transverse
momentum. Therefore the mean transverse momentum of inclusive hadrons
should rise as function of Feynman $x_F$. This effect is indeed observed
in data \cite{seagull} and it allows to measure the jet transverse
momentum which results from the momentum transfer gained in the
interaction and from the primordial intrinsic momentum of the parton. The
latter can be extracted from data and it turns out to be rather high,
$0.5-1\GeV$, the value anticipated in the model of gluonic spots.

\subsection{Unintegrated gluon distribution}

Direct information about transverse momenta of gluons in a hadron comes
from the unintegrated distribution function. There are few
phenomenological distributions in the literature. A quite popular one
\cite{gbw} employs the saturated shape of the dipole cross section fitted
to HERA data for the proton structure function at small Bjorken $x$,
 \beq
 {\cal F}(x,k_T^2) =
\frac{3\,\sigma_0\,R_0^2(x)\,k_T^4}
{16\,\pi^2\,\alpha_s(k_T^2)}\
{\rm exp}\Bigl[-{1\over4}\,R_0^2(x)\,k_T^2\Bigr]\ ,
 \label{700}
 \eeq
 where $R_0(x)=0.4\fm\times(x/x_0)^{0.144}$; $x_0=0.0003$.  From this
distribution the mean transverse momentum squared is $\la
k_T^2\ra=12/R_0^2(x)$. For instance, at $x=0.01$ this is
$\sqrt{\la k_T^2\ra}\approx1\GeV$. Although the distribution
Eq.~(\ref{700}) falls off steeply at large $k_T$, this does not
have a big effect on its mean value. Notice that for the small
$k_T$ region an energy dependent saturated shape of the dipole
cross section is more appropriate \cite{kst2}, which results in a
smaller value for the mean transverse momentum of gluons,
$\sqrt{\la k_T^2\ra}\approx0.6\GeV$ \cite{jkt}.

Both estimates of the mean transverse momentum are quite large indicating
at localization of gluons within small spots.

\subsection{\boldmath$\sigma_L/\sigma_T$ in DIS}

Of course, not only gluons get increased transverse momenta, but also valence
quarks if they are probed with sufficient resolution, $Q^2\gg1/r_0^2$. The
ratio of the longitudinal-to-transverse DIS cross sections,
$R=\sigma_L/\sigma_T$, is sensitive to the transverse motion of valence quarks
(at large $x$), since otherwise is vanishes in LO at high energy $\nu$ (or
high $Q^2$ and fixed $x$) as $R=g(x)Q^2/\nu^2$ according to the Callan-Gross
relation \cite{cg,gj}.

In order to incorporate the effect of transverse motion of the
quarks, a parametrization $R=cQ^2/(Q^2+d^2)^2$ was used in
\cite{bodek} to fit data, with $d^2=0.99\pm0.39\GeV^2$. This
result agrees with the expected $d\sim1/r_0$.

Notice that data for $R_x,Q^2$ is well explained in the dipole
approach \cite{krt} with a saturated cross section. This is not a
surprise, since according to our discussion in the previous
section the dipole cross section bears information about enhanced
gluon momenta.

\bigskip

{\bf Acknowledgments:} We are thankful to Hans-J\"urgen Pirner for helpful
discussions.  This work was supported in part by Fondecyt (Chile) grants
numbers 1030355 and 1050519, and by DFG (Germany) grant PI182/3-1.

\end{document}